\documentclass[aps,prl,twocolumn,superscriptaddress, nofootinbib]{revtex4}

\usepackage[normalem]{ulem}

\usepackage{amsmath}
\usepackage{amsfonts,amssymb,amsthm, bbm,braket}
\usepackage{graphicx}   % need for figures
\usepackage{subfigure}  % and subfigures

\newcommand\abs[1]{\left|#1\right|}

\newcommand\round{\operatorname{round}}

\newcommand\argmin{\operatorname{argmin}}

\newcommand\doi[1]{\href{http://dx.doi.org/#1}{doi:#1}}
\newcommand\arxivref[1]{\href{http://arxiv.org/abs/#1}{arXiv:#1}}

\global\def \arxivmode {}
\ifx \arxivmode \undefined
\else
  \newcommand\refappendix[1]{Appendix #1}
  \newcommand\arxivonly[1]{#1}
  \newcommand\prlonly[1]{}
\fi

\ifx \prlmode \undefined
\else
  \newcommand\refappendix[1]{Appendix #1 of \ref{arxiv-version}}
  \newcommand\arxivonly[1]{}
  \newcommand\prlonly[1]{#1}
\fi

\usepackage{hyperref} % must be loaded last

\begin{document}

%---------------------------------------------------------------------------------------------------------------------
\title{How to best sample a periodic probability distribution, or on the accuracy of Hamiltonian finding strategies}
\author{Christopher Ferrie}
\affiliation{
Institute for Quantum Computing,
University of Waterloo,
Waterloo, Ontario, Canada}
\affiliation{
Department of Applied Mathematics,
University of Waterloo,
Waterloo, Ontario, Canada}

\author{Christopher E. Granade}
\affiliation{
Institute for Quantum Computing,
University of Waterloo,
Waterloo, Ontario, Canada}
\affiliation{
Department of Physics,
University of Waterloo,
Waterloo, Ontario, Canada}

\author{D.G. Cory}
\affiliation{
Institute for Quantum Computing,
University of Waterloo,
Waterloo, Ontario, Canada}
\affiliation{
Department of Chemistry,
University of Waterloo,
Waterloo, Ontario, Canada}
\affiliation{
Perimeter Institute for Theoretical Physics,
Waterloo, Ontario, Canada}

\date{\today}

%- ABSTRACT -----------------------------------------------------------------------------------------------------------

\begin{abstract}
  Projective measurements of a single two-level quantum mechanical system (a qubit) evolving under a time-independent Hamiltonian produce a probability distribution that is periodic in the evolution time. The period of this distribution is an important parameter in the Hamiltonian. Here, we explore how to design experiments so as to minimize error in the estimation of this parameter. While it has been shown that useful results may be obtained by minimizing the risk incurred by each experiment, such an approach is computationally intractable in general. Here, we motivate and derive heuristic strategies for experiment design that enjoy the same exponential scaling as fully optimized strategies. We then discuss generalizations to the case of finite relaxation times, $T_2 < \infty$.
\end{abstract}

\maketitle

%%---------------------------------------------------------------------------------------------------------------------
%\section{Introduction}
%%---------------------------------------------------------------------------------------------------------------------

\emph{Introduction.}  Measurement adaptive tomography has recently been suggested as an efficient means of performing partial quantum process tomography \cite{Sergeevich2011Characterization, ferrie_adaptive_2011}.  Little is known about optimal protocols when realistic experimental restrictions are imposed --- as opposed to the case where one is allowed arbitrary quantum resources\footnote{As in the standard phase estimation protocol. See e.g. \cite{Childs2000Quantum}.}.  Indeed, even in the simplest examples, not even bounds have been given on the proposed protocols.  Here, we give analytic bounds on both non-adaptive and adaptive estimation protocols for a Hamiltonian parameter estimation problem.  Moreover, we derive estimation protocols which asymptotically achieve these bounds.  Adaptive protocols are typically difficult to implement because a complex optimization problem must be solved after each measurement.  We instead derive a heuristic that is easy to implement \emph{and} achieves the exponentially improved asymptotic risk scaling of the optimal solution.

Within the nuclear magnetic resonance (NMR) community, similar concerns have motivated the examination of the use of maximum entropy \cite{barna_exponential_1987} and maximum likelihood \cite{chylla_theory_1995} methods for obtaining spectra. Recently, computational power has become available such as to make these methods feasible for use in analyzing non-uniform data obtained from high-dimensional NMR experiments \cite{hyberts_applications_2011}. These studies have produced qualitatively similar strategies for how to best design experiments when each sample is expensive to collect.

The paper is organized as follows.  First, we define the model Hamiltonian which we want to estimate the parameters of, along with our metric of success.  Then we give both frequentist and Bayesian lower bounds on the \emph{risk} derived from this metric.  Finally, we derive strategies which achieve the asymptotic scaling of these bounds.

\emph{Problem statement.}  The model we consider is a qubit evolving under the Hamiltonian
\[
H = \frac\omega 2\sigma_z.
\]
Here $\omega$ is the unknown parameter whose value we want to ascertain.  We make the problem dimensionless by assuming $\omega\in(0,1)$.  An experiment consists of preparing a single known input state $\ket{+}$, evolving under the Hamiltonian $H$ for a controllable time $t$ and performing a measurement in the $\sigma_x$ basis.  We emphasize here that we are assuming strong projective measurements on individual copies of a quantum preparation, rather than weak measurements on physical ensembles such as those studied in NMR experiments.

The outcomes of the measurement we label $d\in\{0,1\}$, where $0$ and $1$ refer to $\ket+$ and $\ket-$, respectively.  An experiment design consists of a specification of the time $t$ that we evolve a qubit under $H$ before we measure.  The likelihood function for a given experiment $t$ is then given by the Born rule $\Pr(0|\omega,t)=\left\vert\braket{+ | e^{-i H t} | +} \right\vert^2$ and $\Pr(1|\omega,t)=1-\Pr(0|\omega,t)$.  Using our model Hamiltonian, we can express the likelihood more simply as:
\begin{equation}
\label{eq:likelihood-no-t2}
\Pr(d|\omega,t) =  \sin^2\left(\frac\omega 2t\right)^d\cos^2\left(\frac\omega 2t\right)^{1-d}.
\end{equation}
Note that this model does not include noise.  Below, we somewhat generalize this model by including limited visibility and a $T_2$ dephasing process.

If we desire an estimate $\hat\omega$ of the true value $\omega$, a commonly used figure of merit is the \emph{squared error loss}:
\[
L(\omega,\hat \omega) = \abs{\omega-\hat\omega}^2.
\]
The \emph{risk} of an estimator, which is a function that takes data sets $(D,T):=(\{d_k\},\{t_k\})$ to estimates $\hat\omega(D,T)$, is its expected performance with respect to the loss function:
\[
R(\omega,\hat\omega) = \sum_{D} \Pr(D|\omega,T) L(\omega,\hat\omega(D,T)).
\]
For squared error loss, the risk is also called the \emph{mean squared error} (MSE).

%%---------------------------------------------------------------------------------------------------------------------
%\section{Cramer-Rao Lower Bound}
%%---------------------------------------------------------------------------------------------------------------------

\emph{Mean squared error lower bound.}  The difficulty here is that the random outcomes of the measurements are \emph{not} identically distributed.  In fact, since they depend on the measurement time, each one could be different.  Although, asymptotic results exist for non-identically distributed random variables\footnote{The frequentist reference is \cite{Hoadley1971Asymptotic}, while a useful Bayesian reference is \cite{Weng2010Bayesian}.}, these results are derived for insufficient statistics, such as the sample mean.  Moreover, we desire to provide computationally tractable heuristics that permit useful estimates with a finite number of samples.

Although it is quite difficult to obtain exact expressions for the risk for arbitrary measurement times, in some cases we have obtained an asymptotically tight lower bound.  For \emph{unbiased} estimators, we can appeal to the Cramer-Rao bound \cite{Cover2006Elements}
\begin{equation}
R(\omega,\hat\omega) \geq \frac{1}{\mathcal{I}(\omega)}, \label{eq:cramer-rao-bound}
\end{equation}
where
\begin{equation}
\mathcal I(\omega) =  - \sum_{D} \Pr(D|\omega,T) \frac{\partial^2 \log(\Pr(D|\omega,T))}{\partial \omega^2} \label{eq:fisher-def}
\end{equation}
is called the Fisher information. In our particular case, the Fisher information reduces to quite a simple form in
\begin{equation}
\mathcal I(\omega) = \sum_{k=1}^N t_k^2, \label{eq:simple-fisher}
\end{equation}
which is conveniently independent of $\omega$ (a derivation is given in \refappendix{A}).  Thus, the mean squared error is lower bounded by
\begin{equation}\label{eq:freqCRlb}
R(\omega,\hat\omega)\geq\frac{1}{ \sum_{k=1}^N t_k^2}.
\end{equation}
Later we show that this bound becomes exponentially suppressed when we include noise in our model.  In general, this quantity is dependent on the true parameter $\omega$.

The Bayesian solution considers the average of the risk, called the \emph{Bayes risk}, with respect to some prior $\pi(\omega)$:
\[
r(\pi,\hat\omega) = \int R(\omega,\hat\omega) \pi(\omega) d\omega.
\]
As in references \cite{ferrie_adaptive_2011,Sergeevich2011Characterization}, we choose a uniform prior for $\omega\in(0,1)$.  Then, the final figure of merit is the \emph{average} mean squared error:
\[
r(\hat \omega) = \int R(\omega,\hat\omega) d\omega.
\]
The goal is to find a strategy which minimizes this quantity.  Although there exist Bayesian generalizations of the Cramer-Rao bound \cite{Gill1995Applications}, ours is independent of $\omega$ and thus remains unchanged by integrating equation \eqref{eq:freqCRlb} over the parameter space:
\begin{equation}\label{bayesrisklb}
r(\hat \omega) \geq \frac{1}{ \sum_{k=1}^N t_k^2}.
\end{equation}
Note also that, in general, Bayesian Cramer-Rao bounds require fewer assumptions to derive than the standard (frequentist) bound.  Although they are the same for this model, they differ for a more general model considered later.  In broad strokes, the difference in practice between Bayesian and frequentist methods is averaging versus optimization.  Below we demonstrate a heuristic strategy which draws from both methods to achieve the goal of determining the measurement times which give the lowest possible achievable bound on the Bayes risk \eqref{bayesrisklb}.

%% CR BOUND COUNTEREXAMPLE %%
\emph{Looseness of the Cramer-Rao bound.}
As useful as the Bayesian Cramer-Rao lower bound \eqref{bayesrisklb} is, it is simple to see that it is not always achievable.
%To see this, we derive another lower bound by considering the example of a square-wave likelihood function $\Pr(d | \omega, t) = f(\omega t/ 2)$, where $f$ is a function of period $2\pi$ and whose behavior on the interval $[0, 2\pi]$ is given by
%$$
%	f(\theta) = \begin{cases}
%	    0 & 0 \le \theta < \pi \\
%	    1 & \pi \le \theta < 2 \pi
%	\end{cases}.
%$$
%If a system following this likelihood function is sampled at $t=\pi$, then the result gives with perfect certainty whether $\omega < 1/2$ or whether $\omega \ge 1/2$. A measurement at $t = 2\pi$ then divides by half the space of possible parameters, giving whether $\omega \in [0,1/4) \cup [1/2, 3/4)$.
%We can express this by saying that a measurement at $t_k = 2^{k - 1} \pi$ results in the $k^{\mbox{th}}$ digit of the binary expansion of $\omega$ being learned.
%Note that this example gives us a lower bound,
We can obtain a lower bound by considering the best protocol we could possibly hope for in any two-outcome experiment.  In such a protocol, one bit of experimental data provides exactly one bit of certainty about the parameter $\omega$. If we learn the bits of $\omega$ in sequence, at each step $k$, our risk is upper bounded by the worst-case where all the remaining bits of $\omega$ are either all 0 or all 1. In either case, the error incurred by estimating a point between the two extremes is given by $\sum_{n = k+2}^{\infty} 2^{-n} = 2^{-(k+1)}$, leading to the best possible MSE after $N$ measurements being $2^{-2(N+1)}$, even though we can make a smaller Cramer-Rao bound by choosing times that grow faster than this exponential function.  Note that this risk is achievable via the standard phase estimation protocol \cite{Childs2000Quantum}, but that this protocol requires quantum resources which are \emph{not} part of our model.

\emph{Examples.}  Let us consider a couple of examples for which the lower bound can be further simplified.  First, consider the case when all the measurement times are the same.  This is by far the simplest case, since the outcomes become identically distributed.  Recall $\omega\in(0,1)$.  Then, the measurement time should be less then the first Nyquist time, $t\leq\pi$, or the data will be consistent with more than one $\omega$.  That is, for $t>\pi$ (but less than $2\pi$, say), the likelihood function will have two equally likely maxima.  We minimize the risk, then, by choosing $t=\pi$.  Then, the maximum likelihood estimator (MLE), for example, will be asymptotically efficient \cite{Lehmann1998Theory} achieving the Cramer-Rao lower bound
\[
r(\hat\omega_{\rm MLE}) = \frac{2}{\pi^2 N} +O\left(\frac 1{N^2}\right).
\]

Now consider a uniform grid of times.  Since $\omega\in(0,1)$, we should choose the Nyquist sampling rate: $t_k = k\pi$.  Then, for any estimator $\hat{\omega}$ using data collected at these measurement times, the Cramer-Rao bound gives
\[
r(\hat\omega)\geq\frac{6}{\pi^2 N (1+N)(1+2N)}=\frac{3}{\pi^2 N^3}+O\left(\frac1{N^4}\right).
\]
Again, the maximum likelihood estimator will be asymptotically efficient.  However, since the likelihood function will have many local maxima, the maximum likelihood estimator is non-trivial to find as gradient methods are not guaranteed to work.  Bayesian estimators were derived in \cite{Sergeevich2011Characterization}, where simulations yielded $\sim1/N^3$ risk scaling which is asymptotically efficient.

Note that since we are considering a uniform spacing of times, we can apply a Fourier estimation technique without worrying about spectral aliasing introduced by non-uniformity \cite{maciejewski_nonuniform_2009}.  That is, we apply the discrete Fourier transform and estimate the peak of the power spectrum.  Since the resolution in the frequency domain is $1/N\triangle t$, we expect the Bayes risk to be
\[
r(\hat\omega_{\rm Fourier}) = \frac{1}{\pi^2 N^2}.
\]
The sampling theorem requires that we sample from a \emph{deterministic} function, not a probability distribution.  In practice, this condition is often approximately satisfied by sampling some stable statistic such as the mean value of the distribution at each time.  This can be achieved by measuring at the same time until a sufficiently accurate estimate of the mean at that time is obtained, then repeating this for many other times.  But as we have shown, this method can be quadratically improved by performing every \emph{single} measurement at a different time.

\emph{Exponentially achievable lower bound.}  It has been shown that Bayesian adaptive solutions lead to risk decreasing exponentially with the number of measurements \cite{Sergeevich2011Characterization}.  However, these results are given by fits to numerical data.  Here, we give an analytic lower bound on the risk of these protocols.

The local (in time) Bayesian adaptive protocol can be described as follows: (1) begin with a uniform prior $\Pr(\omega)$ and determine the first measurement time $t_1\approx1.136\pi$ which minimizes the average (over the two possible outcomes) variance of the posterior distribution; (2) perform a measurement at $t_1$, record the outcome $d_1$, and update the distribution $\Pr(\omega)\mapsto \Pr(\omega|d_1,t_1)$ via Bayes' rule; (3) repeat step (1) replacing the current prior with the current posterior.  Note that the expected variance in the posterior is the Bayes risk.  Thus, the protocol attempts to minimize the risk assuming the next measurement is the last. Strategies that are local in this sense are called a \emph{greedy} strategies, as opposed to strategies which attempt to minimize the risk over all future experiments.

For some choices of measurement times, including those given by the protocol above, the posterior will be approximately normally distributed\footnote{This is true asymptotically and higher order corrections can be used if required \cite{Weng2010Bayesian}.}.  This is guaranteed in the asymptotic limit, but the posterior distribution near its peak is also remarkably well approximated by a Gaussian after as few as 15 reasonably chosen measurements (we found a uniform grid $t_k=k\pi$ to be sufficient for ``warming up'' to the Gaussian approximation).  Thus, we approximate the current distribution (at given some sufficiently long measurement record $D$) as
\[
\Pr(\omega|D) = \frac{1}{\sqrt{2 \pi \sigma^2 }}e^{-\frac{(\omega-\mu )^2}{2 \sigma^2 }},
\]
with some arbitrary mean $\mu$ and variance $\sigma^2$ implied by $D$.  The expected posterior variance (which is equal to the Bayes risk) of the probability distribution of the next measurement is
\begin{equation}
r(t) = \sigma^2\left(1 +\frac{ t^2 \sigma^2 \sin(\mu t)^2}{- e^{t^2 \sigma^2 }+\cos(\mu t)^2}\right), \label{eq:bayes-risk-next}
\end{equation}
(derived in \refappendix{B}) which oscillates with frequency $2\mu$ within an envelope $\sigma^2 \left(1- t^2 \sigma^2  e^{-t^2 \sigma }\right)$.  Asymptotically, the minimum risk will approach the minimum of the envelope for all $\mu$, but will be a lower bound on the risk otherwise.  This minimum occurs at $\hat t=\frac1\sigma$ with a risk of $r(\hat t) = (1 -e^{-1})\sigma^2$, which is also the variance of the updated probability distribution since both outcomes are equally probable at $\hat t$.  Thus, at each measurement step we reduce the risk by $1-e^{-1}\approx0.632\approx e^{-0.459}\approx 2^{-0.661}$.  Thus, the risk scales exponentially as $r\sim \sigma^2 (1-e^{-1})^N$ and is achieved at measurement times which scale as
\[
t_k\sim\frac1{\sigma(1-e^{-1})^{k/2}}\approx \frac{1.26^k}\sigma.
\]

These times are guaranteed to be optimal only in the asymptotic limit.  For finite numbers of samples, we suggest two simple heuristics.  First, we suggest the use of exponentially increasing times, where the base of the exponent is optimized offline, followed by the use of the maximum likelihood estimator for these times.  Second, we suggest a simpler adaptive scheme based on the assumption that the distribution remains Gaussian after each measurement.  Making use of this normality assumption, we only need update equations for the mean and variance of the distribution over $\omega$. In deriving the update equations, we also take into account the oscillations of the expected Bayes risk by finding the nearest achievable minima to the one given by the lower bound.  We provide the update equations in \refappendix{C}.

\begin{figure}
  \includegraphics[width=\columnwidth]{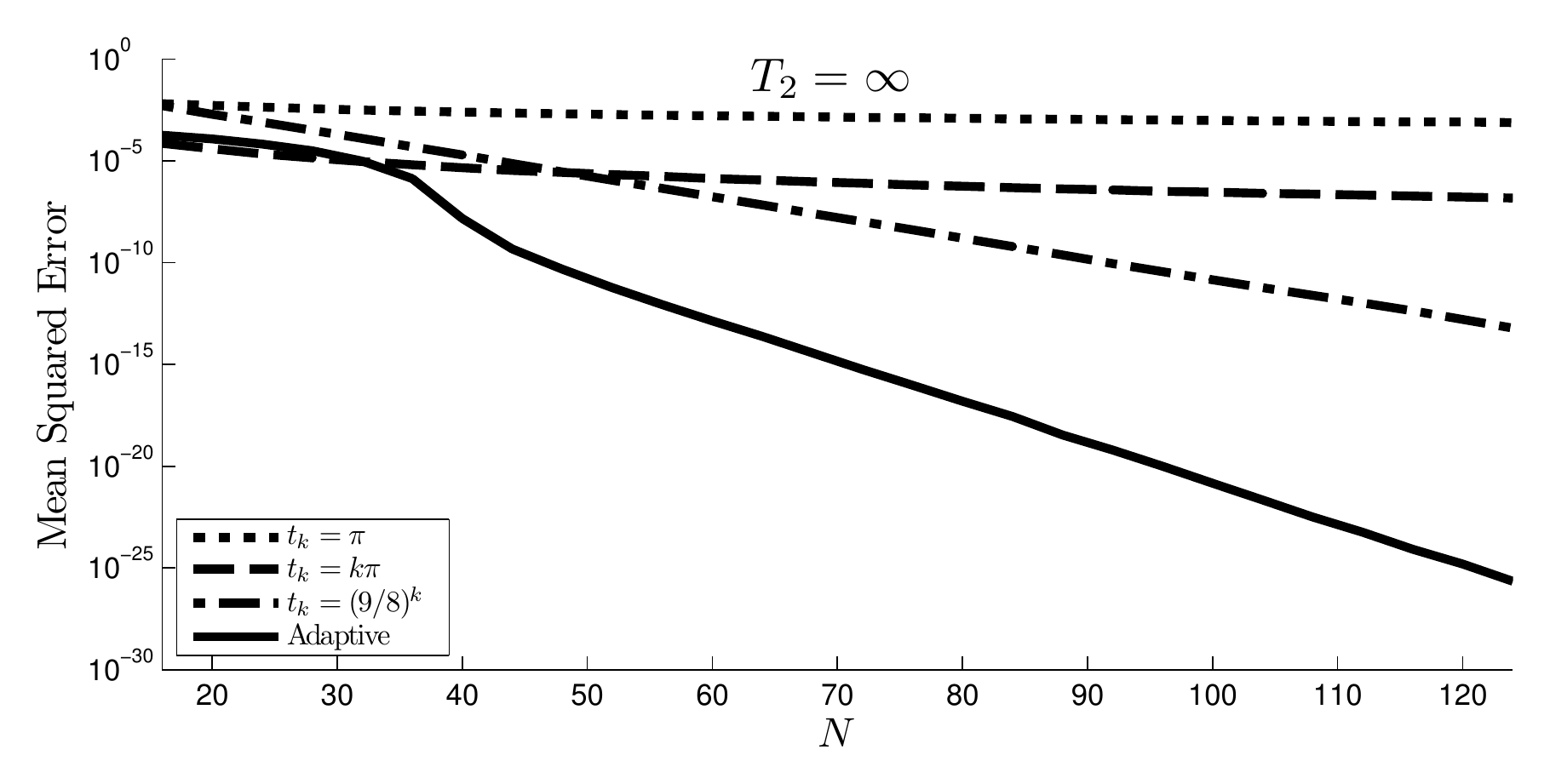}\\
  \includegraphics[width=\columnwidth]{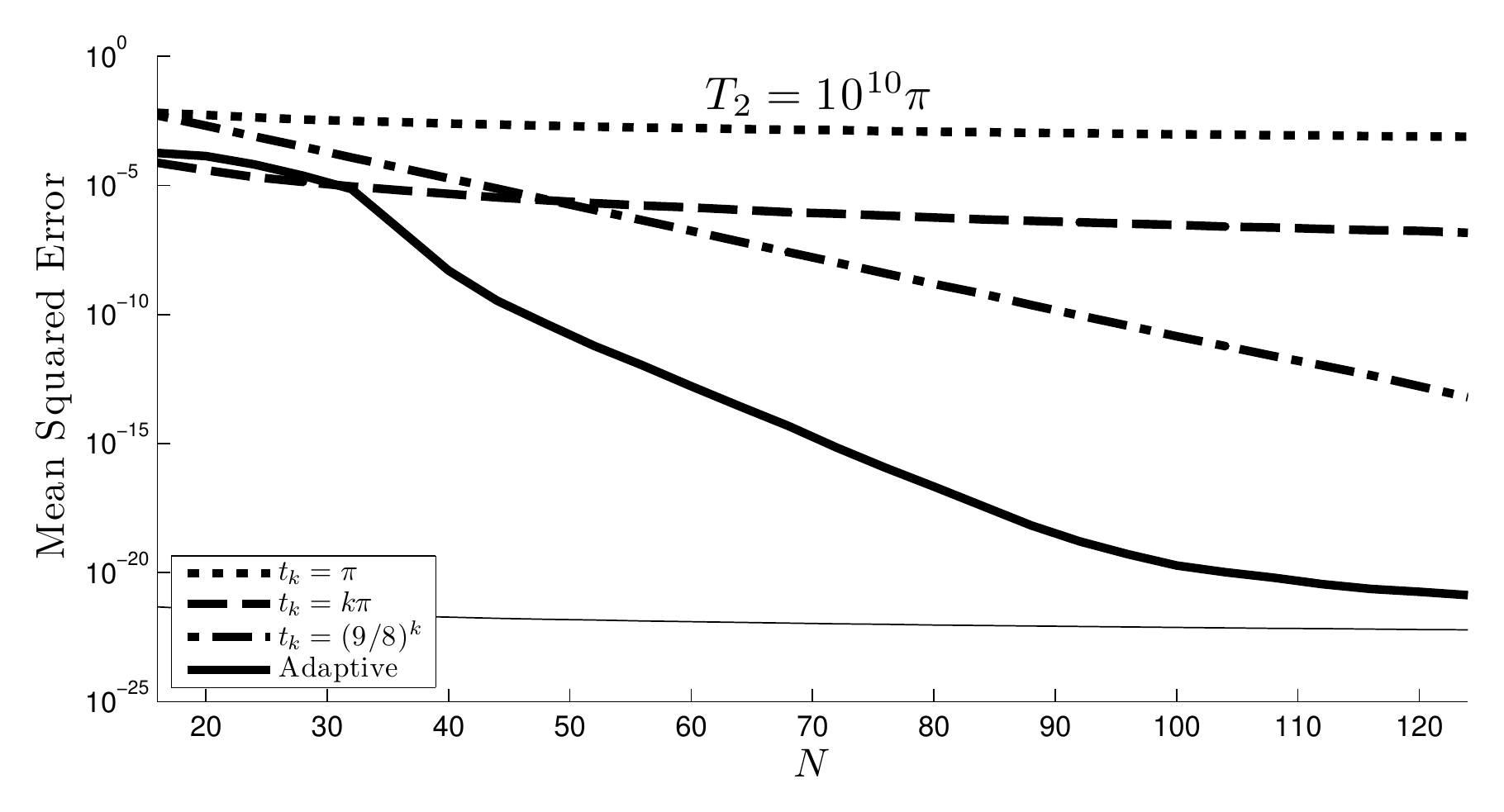}\\
  \includegraphics[width=\columnwidth]{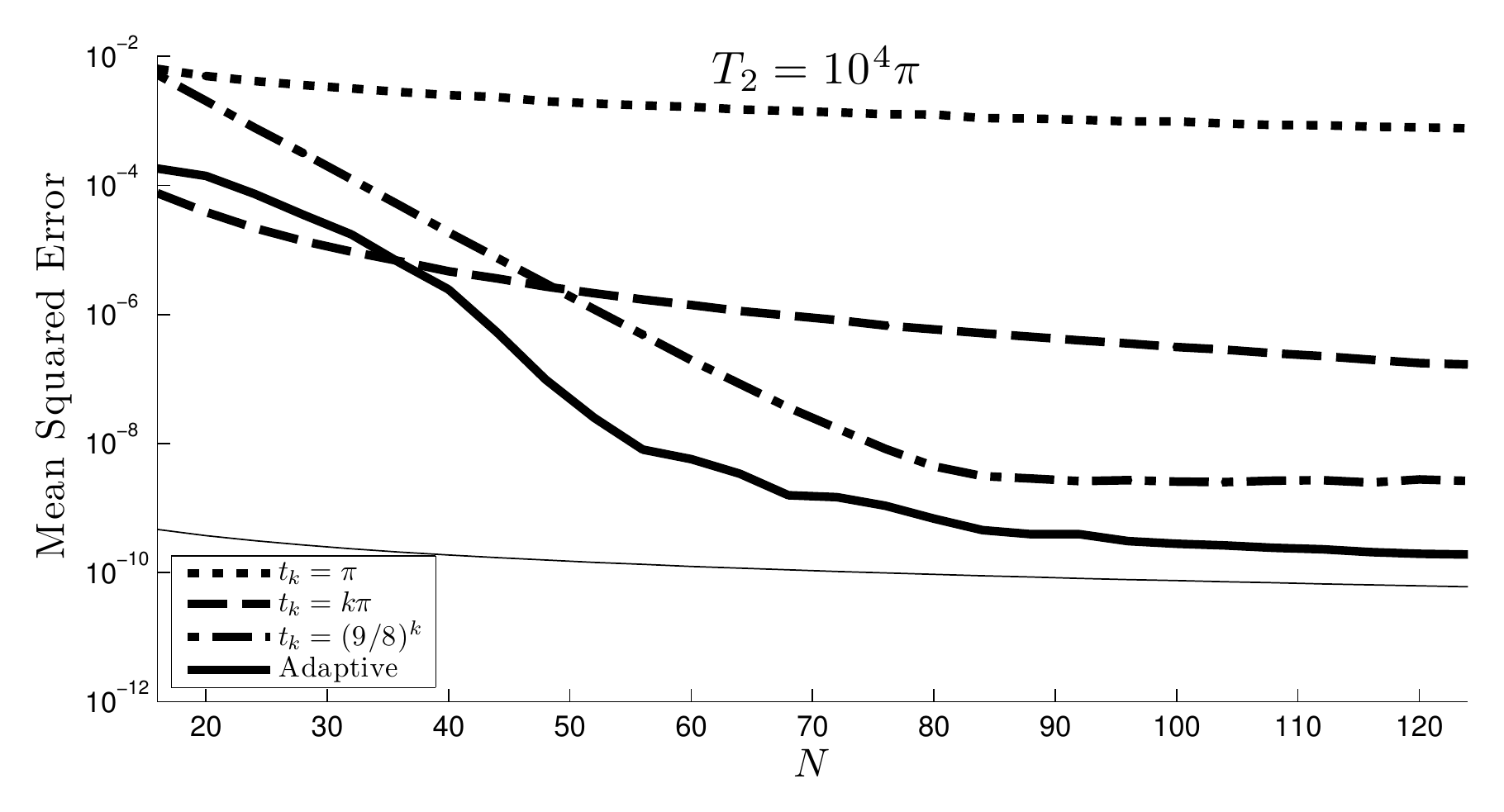}
  \caption{\label{fig:results} The Bayes risk -- the average (over a uniform prior) mean (over data) squared error -- of the strategies discussed in the paper.  Data points are at evenly spaced measurement numbers $N\in\{16,20,24,\ldots,124\}$ and the lines are linear interpolants to guide the eye.  Each data point is the average of $10^4$ simulations.  In each figure, the noise parameter $\eta=1$ since its inclusion only gives a constant offset.  From top to bottom, the relaxation characteristic time is $T_2=\infty, 10^{10} \pi, 10^4 \pi$. The thin solid lines indicate the lower bound given by Equation \eqref{eq:ultimate-lower-bound}.}
\end{figure}

\emph{Generalization to finite $T_2$.}  In practice, we will have to consider not only experimental restrictions but also noise and relaxation processes.  Processes which do not affect the quantum state can be effectively modeled by random bit-flip errors occurring with probability $1-\eta$.  Processes which do affect the quantum state (decoherence) are modeled by an exponential decay of phase coherence\footnote{We do not include amplitude damping in our model since our populations remain equal throughout evolution and thus $T_1$ only manifests as a contribution to $T_2$.} with characteristic time $T_2$. Since the state being measured lies in the $xy$-plane of the Bloch sphere, this loss of phase coherence manifests as an exponential decaying envelope being applied to the original likelihood (\ref{eq:likelihood-no-t2}). The model is thus fully specified by the likelihood function
\begin{equation}
\begin{split}
\label{eq:likelihood-finite-t2}
\Pr(0|\omega,t,\eta,T_2) = \mspace{-96mu} & \\ & \eta\left(e^{-\frac{t}{T_2}}\cos^2\left(\frac\omega 2 t\right)+\frac{1-e^{-\frac{t}{T_2}}}2\right)+\frac{1-\eta}2.
\end{split}
\end{equation}

The Cramer-Rao bound is now given by
\begin{equation}
R(\omega,\hat\omega)\geq \left(\sum_{k=1}^N \frac{t^2_k\eta^2\sin^2(\omega t_k)}{e^{\frac{2t_k}{T_2}}-\eta^2\cos^2(\omega t_k)}\right)^{-1}. \label{eq:cramer-rao-bound-finite-T2}
\end{equation}
Note that unlike the Cramer-Rao bound \eqref{eq:freqCRlb} for the noiseless case, the above bound is not independent of $\omega$ and thus we must appeal to the Bayesian Cramer-Rao bound so that the measurement times can be chosen independently of the true parameter.  However, the Bayesian bound turns out to be very loose.  A sharper bound is given by first upper bounding each term in the denominator to give
\[
r(\hat\omega)\geq\frac{1}{\eta^2\sum_{k=1}^N t^2_k e^{-\frac{2t_k}{T_2}}}.
\]
The noise term (or visibility) $\eta$ simply gives a constant reduction in the achievable accuracy.  The relaxation process provides a more interesting dynamic as we see that the gains from longer times are exponentially suppressed.  In other words, strategies are restricted to explore $t_k\leq T_2$.  We can thus do no better than
\begin{equation}
r(\hat\omega)\geq\frac{e^{2}}{N\eta^2T^2_2}. \label{eq:ultimate-lower-bound}
\end{equation}

The adaptive strategy discussed above can be generalized to include noise and relaxation but the expressions are more lengthy (see \refappendix{B}).  To illustrate the performance of our adaptive strategy, we simulate the adaptive strategy along with offline strategies using identical times ($t_k = \pi$), linearly spaced times ($t_k = k\pi$) and exponentially sparse times ($t_k = (9/8)^k$). For each strategy, we perform simulations for experiments consisting of different numbers of samples $N$, up to $N = 124$, and repeat each such simulation $10^4$ to obtain an estimate of the Bayes risk for that strategy and experiment size. In Fig. \ref{fig:results}, we present the results of these simulations for the noiseless case, and for the cases $T_2 = 10^{10} \pi$ and $T_2 = 10^4 \pi$.

Note that in all cases, the adaptive strategy achieves exponential scaling until the times selected reach $t=T_2$.  At that point, the risk will then scale linearly if the remaining measurement times are $t=T_2$.  However, if the protocol continues to select larger measurement times, the information gained from those measurements will tend to zero and the risk will remain constant.

\emph{Summary and conclusions.} By using the Cramer-Rao bound along with analytic expressions for the variance of each posterior distribution, we have motivated a heuristic method for choosing experiment designs that asymptotically admits exponentially small error scaling in the number of measurements. For finite measurements, we have relied on numerical simulation to demonstrate that this scaling is well-achieved even for $N \lessapprox 120$. Numerical simulations for finite $T_2$, moreover, have suggested that we can enjoy exponential scaling of the risk until the measurement times saturate the $T_2$ bound, at which point the risk scaling switches to the asymptotic scaling of $1/N$. In both cases, the heuristics used to design experiments are quite computationally tractable, thus motivating the utility of our heuristics to actual experimental practice.

\emph{Acknowledgements.}
We thank Miriam Diamond for assistance in testing and developing the simulation software.  CF thanks Josh Combes for helpful discussions.  This work was financially supported by NSERC and CERC.

%\arxivonly{
\onecolumngrid
\appendix

\section{Appendix A: Derivation of Cramer-Rao Bounds}
In this Appendix, we show that for the simple model represented by the likelihood function presented in equation \eqref{eq:likelihood-no-t2}, the Fisher information given by \eqref{eq:fisher-def} reduces to the form claimed in \eqref{eq:simple-fisher}. To show this, we first note that the likelihood for a vector $D = (d_1, d_2, \dots, d_k)$ of observations at times $T = (t_1, t_2, \dots, t_k)$ is given by a product of the likelihoods for each individual measurement,
\[
  \Pr(D | \omega, T) = \prod_k \Pr(d_k | \omega, t_k).
\]
Thus, the log-likelihood function is simply a sum over the individual log-likelihoods. Since the derivative operator commutes with summation, we obtain that
\[
 \frac{\partial^2}{\partial\omega^2} \log \Pr(D | \omega, T) = \sum_k \frac{\partial^2}{\partial\omega^2} \log \Pr(d_k | \omega, t_k).
\]
This in turn implies that the Fisher information for a vector of measurements is given by the sum for each measurement of that measurement's Fisher information.

To calculate the single-measurement Fisher information, we find the second derivative of the log-likelihood for a single measurement is given by
\[
  \frac{\partial^2}{\partial\omega^2} \log \Pr(d_k | \omega, t_k) = t_k^2 \frac{\left(2 d_k-1\right) \left(1-2 d_k+\cos \left(\omega  t_k\right)\right)}{\left(\left(2 d_k-1\right) \cos \left(\omega  t_k\right)-1\right)^2}.
\]
Thus, we find that the single-measurement Fisher information is given by
\begin{align*}
 \mathcal{I}(\omega | t_k) & = -\sum_{d_k\in\{0,1\}} \Pr(d_k | \omega, t_k) \frac{\partial^2}{\partial\omega^2} \log \Pr(d_k | \omega, t_k) \\
 & = t_k^2 \sum_{d_k\in\{0,1\}} \frac{\left(2 d_k-1\right)\left(1-2 d_k+\cos \left(\omega  t_k\right)\right)}{2 \left(2 d_k-1\right) \cos \left(\omega  t_k\right)-2} \\
 & = t_k^2.
\end{align*}
We conclude that $\mathcal{I}(\omega | T) = \sum_k t_k^2$, as claimed.

For the model with finite $T_2$ and limited visibility, given by the likelihood function \eqref{eq:likelihood-finite-t2}, we can follow the same logic. We find the second derivative of \eqref{eq:likelihood-finite-t2} with respect to $\omega$ gives us
\[
 \frac{\partial^2}{\partial\omega^2} \log \Pr(d_k | \omega, t_k)
  = \eta t_k^2 \cdot \frac{\left(2 d_k-1\right) \left(\eta\left(1 -2 d_k\right) +e^{\frac{t_k}{T_2}} \cos \left(\omega  t_k\right)\right)}{\left(\eta\left(1 -2 d_k\right) \cos \left(\omega  t_k\right)+e^{\frac{t_k}{T_2}}\right)^2}
.
\]
The expected value of this derivative then gives us the Fisher information for a single measurement in the finite-$T_2$ model,
\[
 \mathcal{I}(\omega | t_k) = \frac{\eta ^2 t_k^2 \sin ^2\left(\omega  t_k\right)}{e^{\frac{2 t_k}{T_2}}-\eta ^2 \cos ^2\left(\omega  t_k\right)}.
\]
Taking the sum of this information then produces the Cramer-Rao bound given in \eqref{eq:cramer-rao-bound-finite-T2}.

\section{Appendix B: Asymptotic Scaling of the Bayes Risk}
In this Appendix, we derive expressions for posterior distributions under the assumption of a normally-distributed prior, and then apply these expressions to show the asymptotic scaling of the Bayes risk. We also derive update rules that allow for expedient implementation of the greedy algorithm described in the main text.

Under the assumption of a normally-distributed prior, all prior information about the parameter $\omega$ can be characterized by the mean $\mu$ and variance $\sigma^2$ of the prior distribution. Thus, we shall write our priors as $\Pr(\omega | \mu, \sigma^2)$ to reflect the assumption of normality. Then, the probability of obtaining a datum $d$ at time $t$ given such prior information is then given by
\[
 \Pr(d|t; \mu,\sigma^2) = \int_{-\infty}^\infty \Pr(d|t, \omega) \Pr(\omega | \mu, \sigma^2) d\omega
 = \frac{1}{4} \left(2-(2 d-1) \left(1+e^{2 i \mu  t}\right) e^{-\frac{1}{2} t \left(\sigma ^2 t+2 i \mu \right)}\right).
\]
Applying Bayes' rule then produces the posterior distribution
\[
\begin{split}
 \Pr(\omega|d,t;\mu,\sigma^2) & = \frac{ \Pr(\omega|\mu,\sigma^2) \Pr(d|t,\omega) } { \Pr(d|t; \mu,\sigma^2) } \\
 & = \frac{\sqrt{\frac{2}{\pi }} e^{-\frac{(\mu -\omega )^2}{2 \sigma ^2}} ((1-2 d) \cos (t \omega )+1)}{\sigma  \left(2-(2 d-1) \left(1+e^{2 i \mu  t}\right) e^{-\frac{1}{2} t \left(\sigma ^2 t+2 i \mu \right)}\right)}.
\end{split}
\]
The mean and variance of this distribution are given by:
%\begin{widetext}
\begin{align*}
 \mathbb{E}[\omega | d, t; \mu, \sigma^2]  & = \frac{2 \left((2 d-1) e^{-\frac{1}{2} \sigma ^2 t^2} \left(\sigma ^2 t \sin (\mu  t)-\mu  \cos (\mu  t)\right)+\mu \right)}{2-(2 d-1) \left(1+e^{2 i \mu  t}\right) e^{-\frac{1}{2} t \left(\sigma ^2 t+2 i \mu \right)}} \\
 \mathbb{V}[\omega | d, t; \mu, \sigma^2] & = \mu ^2+\sigma ^2-\frac{2 \left((2 d-1) e^{-\frac{1}{2} \sigma ^2 t^2} \left(\sigma ^2 t \sin (\mu  t)-\mu  \cos (\mu  t)\right)+\mu \right)}{2-(2 d-1) \left(1+e^{2 i \mu  t}\right) e^{-\frac{1}{2} t \left(\sigma ^2 t+2 i \mu \right)}}\\
 & \quad -\frac{2 (2 d-1) \sigma ^2 t e^{i \mu  t} \left(\sigma ^2 t \cos (\mu  t)+2 \mu  \sin (\mu  t)\right)}{(2 d-1) \left(1+e^{2 i \mu  t}\right)-2 e^{\frac{1}{2} t \left(\sigma ^2 t+2 i \mu \right)}}
\end{align*}
%\end{widetext}

To chose optimal times, we wish to pick $t$ so as to minimize the expected value over of the variance, where this expectation is taken over possible data. Based on the previous expressions, we find that
\[
  \mathbb{E}_d[\mathbb{V}_\omega[\omega | d, t; \mu, \sigma^2]] = \sigma^2\left(1 +\frac{ t^2 \sigma^2 \sin(\mu t)^2}{- e^{t^2 \sigma^2 }+\cos(\mu t)^2}\right),
\]
in agreement with Equation \eqref{eq:bayes-risk-next}.

\begin{figure}
  \includegraphics[width=.45\columnwidth]{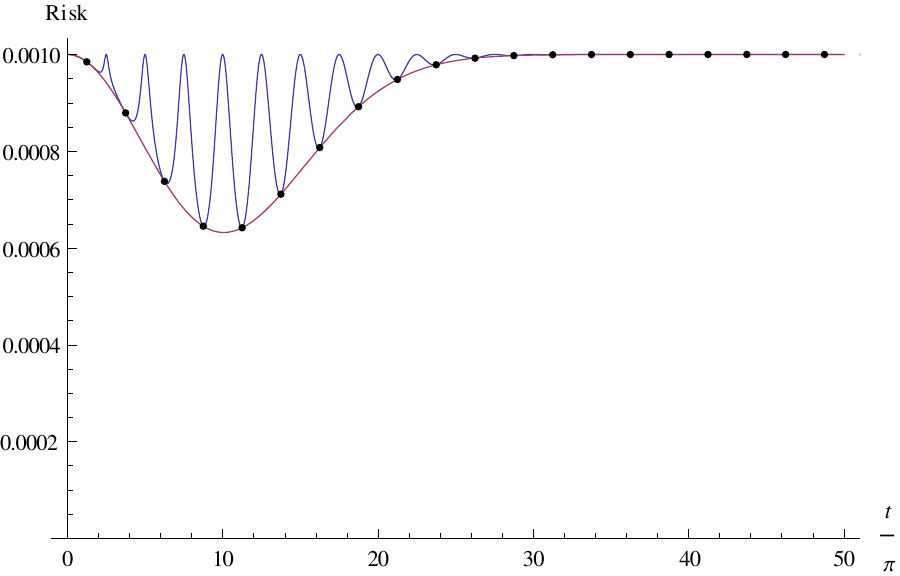}
  \includegraphics[width=.45\columnwidth]{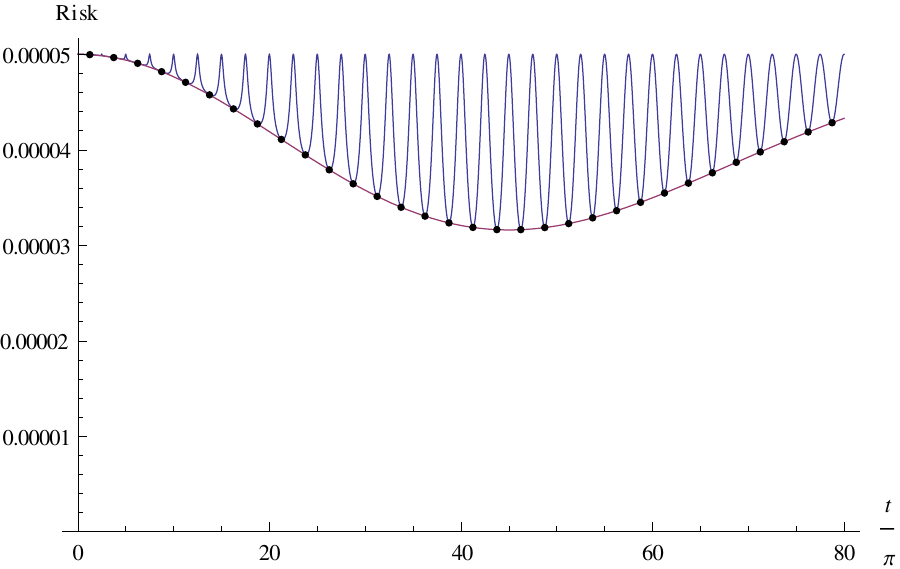}
  \caption{\label{fig:risk-envelope} The risk envelope $E(t, \sigma^2)$, and the risk $r(t; \mu, \sigma^2) \ge E(t, \sigma^2)$ for the examples where $\mu=0.4$ and $\sigma^2 = 10^{-3}$ (left) and $\sigma^2 = 5\times 10^{-5}$ (right). Note that as $\sigma^2$ shrinks, there intersections between $E$ and $r$ (marked by dots) become more tightly packed.}
\end{figure}
This expected variance, which describes our risk incurred by measuring at a given $t$, is bounded below by an envelope $E(t, \sigma^2) = \sigma^2 \left(1-t^2 \sigma^2 e^{-t^2 \sigma^2}\right)$. A pair of examples of the envelope $E(t,\sigma^2)$ and achievable risk $r(t;\mu, \sigma^2)$ is illustrated in Figure \ref{fig:risk-envelope}.

Note that the envelope is minimized by $\hat{t} = \argmin_t E(t, \sigma^2) = 1/\sigma$. Moreover, the expected variance saturates the lower bound at intervals in $t$ of $1/\mu$, but the width of the envelope's minimum grows as $1/\sigma^2$, so that as more measurements are performed, the bound becomes a good approximation for the minimum achievable risk. Thus, in the asymptotic limit of large numbers of experiments, we have that the risk at scales with each step as the minimum of the envelope,
\[
 \frac{E(\hat{t},\sigma^2)}{\sigma^2} = 1-e^{-1} \approx 0.632.
\]
We conclude that in the asymptotic limit, the risk decays as $e^{N \ln 0.632} \approx e^{-0.458 N}$, where $N$ is the number of measurements performed.

\section{Appendix C: Update Equations for $\mu$, $\sigma^2$}

In this Appendix, we state without derivation the update rules for $\mu$ and $\sigma^2$ after obtaining a measurement result $d$ from an experiment performed at time $t$, under the assumption of an normal prior. For the simple model described by Equation \eqref{eq:likelihood-no-t2},
\begin{align}
 \mathbb{E}\left[\omega |d\right] & =\mu -\frac{\pi  (2 d-1) \sigma ^2 (-1)^{k} \left(2 k-1\right) \exp \left(-\frac{\pi ^2 \sigma ^2 \left(1-2 k\right)^2}{8 \mu ^2}\right)}{2 \mu } \\
 \mathbb{V}\left[\omega |d\right] & =\sigma ^2-\frac{\pi ^2 (1-2 d)^2 \sigma ^4 \left(1-2 k\right)^2 \exp \left(-\frac{\pi ^2 \sigma ^2 \left(1-2 k\right)^2}{4 \mu ^2}\right)}{4 \mu ^2},
\end{align}
where $k = \round\left[\frac{\mu }{\pi  \sigma }+\frac{1}{2}\right]$ is used to pick the intersection of $E(t, \sigma^2)$ and $r(t; \mu, \sigma^2)$ to the minimum of $E$, as described in Appendix B.

For the finite-$T_2$ model, the updated mean and variance are given by
\begin{align}
 \mathbb{E}\left[\omega |d\right] & = \mu + \frac{\pi  (2 d-1) (-1)^k (2 k-1) \sigma ^2 \exp \left(-\frac{(\pi -2 \pi  k) \left(-2 \pi  k \sigma ^2 T_2+4 \mu +\pi  \sigma ^2 T_2\right)}{8 \mu ^2 T_2}\right)}{2 \mu } \\
 \mathbb{V}\left[\omega |d\right] & = \sigma ^2-\frac{\pi ^2 (2 d-1)^2 (2 k-1)^2 \sigma ^4 \exp \left(-\frac{(\pi -2 \pi  k) \left(-2 \pi  k \sigma ^2 T_2+4 \mu +\pi  \sigma ^2 T_2\right)}{4 \mu ^2 T_2}\right)}{4 \mu ^2},
\end{align}
where in this case,
\[
 k = \round\left[\frac{\mu -\mu  \sqrt{4 \sigma ^2 T_2^2+1}+\pi  \sigma ^2 T_2}{2 \pi  \sigma ^2 T_2}\right].
\]

%} % end arxivonly.

%\bibliographystyle{plain}
%\bibliography{csferrie}
\twocolumngrid

\end{document}